\documentstyle[epsfig]{aipproc}

\begin{document}
\title{Physics in Ultra-strong Magnetic Fields}

\author{Robert C. Duncan}
\address{Dept.~of Astronomy and McDonald Observatory 
\\University of Texas at Austin}

\maketitle

\begin{abstract}
In magnetic fields stronger than $B_Q \equiv m_e^2 c^3/\hbar e = 4.4
\times 10^{13}$ Gauss, an electron's Landau excitation energy exceeds
its rest energy. I review the physics of this strange regime and some of
its implications for the crusts and magnetospheres of neutron
stars. In particular, I describe how ultra-strong fields
\begin{list}{$\bullet$}{\setlength{\leftmargin}{1.8\leftmargini}
\setlength{\rightmargin}{\leftmargin} \setlength{\itemsep}{0pt}
\setlength{\parsep}{0pt} \setlength{\topsep}{2pt} }
\item render the vacuum {\it birefringent\/} and capable of 
distorting and magnifying images (``magnetic lensing"); 
\item change the self-energy of electrons: as $B$ increases
they are first slightly lighter than $m_e$, then slightly heavier; 
\item cause photons to rapidly split and merge with each other; 
\item distort atoms into long, thin cylinders and molecules 
into strong, polymer-like chains; 
\item enhance the pair density in thermal pair-photon gases; 
\item strongly suppress photon-electron scattering, and
\item drive the vacuum itself unstable, at extremely large $B$. 
\end{list}
In a concluding section, I discuss the spindown of 
ultra-magnetized neutron stars and recent soft gamma repeater
observations. 
\end{abstract}

\section*{Electrons at \boldmath $B> B_Q$ }

The significance of the quantum electrodynamic field strength, $B_Q$,
can be understood via a simple, semi-classical argument. 
A classical electron gyrating in a magnetic field satisfies 
$\dot{p} = ev\times B/c$, where $p= \gamma m_e v$  is the momentum. 
Substituting \hbox{$\dot{p} = \omega p$} and $v = \omega r$ in this equation
and cancelling factors of $\omega$ (along with orbital phase factors), 
one finds a radius of gyration $r = cp/eB$, where $p$ is the transverse
momentum (${\bf \perp B}$). \ 
Quantum mechanics implies $ r \cdot p \sim  \hbar$ in the ground state,  
thus the semi-classical gyration radius is 
\hbox{$r_{\rm gyr} \sim \lambda_e \, (B/B_Q)^{-1/2}$}, where 
\hbox{$\lambda_e \equiv \hbar/m_e c$} is the electron Compton wavelength.
The associated momentum is
\hbox{$p \sim (\hbar/r_{\rm gyr}) \sim  m_e c \, (B/B_Q)^{1/2}$.} \

This shows that electrons gyrate {\it relativistically} in fields
$B > B_Q$.  \ One thus expects excitation energies in excess of $m_e c^2$.
This is borne out by the solution to the Dirac 
equation for an electron in a homogeneous magnetic field.  The Dirac 
spinors are proportional to Hermite polynomials, and the energy levels 
or ``Landau levels"  are \hfil\break
\smallskip
\qquad\qquad\qquad\qquad\qquad
$E_n = \left[ m_e^2 + p_z^2 + m_e^2 \, n 
\, (2 B/ B_Q) \right]^{1/2}, $ \hfil (1)\break
in units with $\hbar = c =1$ (adopted also in many equations 
that follow).  
The first term in the square brackets is the rest energy.
The $p_z$-term gives the energy of motion parallel to the field,
which can take a continuum of
values. The discrete energy levels are given by 
$n = 0, 1, 2 \ldots$ \ These states are also eigenstates of spin,
with the $n = 0$ ground state always having $s = -{1\over2}$. \
For $p_z = 0$, the ground state energy is $E_o = m_e$, independent of
$B$. \ 
In a semi-classical picture, one could say that the negative spin-alignment
energy in the ground state cancels with the zero-point gyration
energy.  Excited Landau levels are two-fold degenerate in $s$. 
\  The first Landau-level excitation energy is
\hbox{$\omega_B(1)  = E_1 - E_o\approx (2B/B_Q)^{1/2} \, m_e$} for $B\gg B_Q$. 
\ Because this energy is so large,
electrons almost always remain in the ground state for 
processes thought to occur near the surfaces of ultra-magnetized
neutron stars.

Electron {\it self-interactions}
resolve the degeneracies of the Landau levels, and shift the ground 
state energy.  This
was first demonstrated by Schwinger, who estimated the ``anomalous"
magnetic moment of the electron \cite{schw48}.  The relevant Feynman
diagram is shown in Figure 1: a free electron (traveling upward on the
page) emits a virtual photon, interacts with the magnetic field, then
reabsorbs the photon.  The electron's effective spin magnetic moment 
is enhanced by $(1 + \alpha/2\pi)$ to first order in $\alpha = e^2/\hbar c =
1/137$, the fine-structure constant.  This results
in a ground-state energy shift\hfil\break
\smallskip
\qquad \qquad \qquad \qquad \qquad$
E_o = m_e \ [ \,1 - (\alpha/2\pi) \, (B/ B_Q)\, ]^{1/2} \ .
$ \hfil (2)\break

\begin{figure}[b!] 
\centerline{\epsfig{file=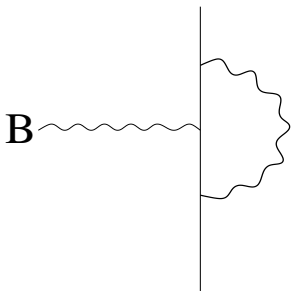,height=1.5in,width=1.5in}}
\vspace{10pt}
\caption{Anomalous magnetic moment diagram.}
\label{fig1}
\end{figure}

If extrapolated to $B > B_Q$, this formula would imply that the
ground-state energy of an electron goes to {\it zero} at 
$B = (2\pi/\alpha)B_Q \approx 4 \times 10^{16}$ Gauss.  
For stronger $B$, the vacuum would become unstable 
to pair production, with dramatic astrophysical consequences\cite{ocon68}.
But eq.~(2) is actually only valid in the sub-$B_Q$ regime.\ 
More generally, the electron's self-energy is determined by 
the sum of Feynman diagrams shown in Figure 2 (ref.~\cite{deme53}). \ 
The triple line on the left-hand side represents 
the physical electron propagator (i.e., the probability amplitude for
an electron to move from point A to point B). 
The double lines on the right are bare propagators for 
an electron in the presence of a magnetic field, 
corresponding to basis states with energies given by
eq.~(1).\footnote{Self-interactions also occur when $B=0$. \ 
The double-line propagators of Fig.~2 are then replaced
with single-line, free-electron propagators (plane-wave states), 
and the resultant energy shifts---formally divergent---are absorbed into 
the electron's known rest mass by the renormalization of quantum 
electrodynamics. A strong magnetic field changes the self-energy when 
the same renormalization prescriptions are used.  
\ Note that the Schwinger 
diagram of Fig.~1 is included in the second diagram on the right of Fig.~2: 
when \hbox{$B < B_Q$}, the double-line propagators can be 
approximated as single-line, free electron propagators 
undergoing discrete, perturbative interactions with the magnetic field. 
Positron intermediate states are included; they correspond to a subset
of the vertex time-orderings which are summed over.
The lowest-order tadpole diagram gives no contribution in a homogeneous 
magnetic field.}

\begin{figure}[b!] 
\centerline{\epsfig{file=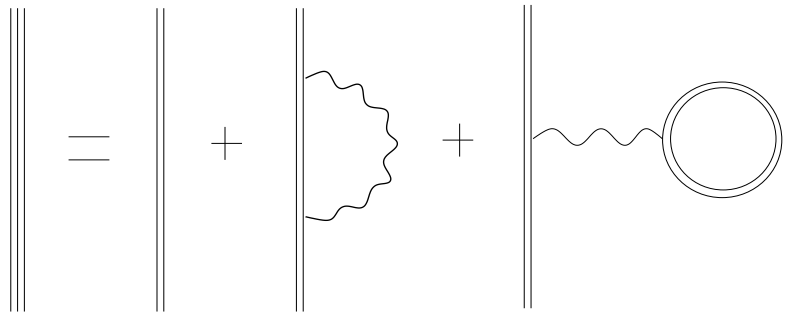,height=1.62in,width=4.05in}}
\vspace{10pt}
\caption{Electron self-energy in a magnetic field: lowest-order diagrams.}
\label{fig2}
\end{figure}

  When the calculation is done, it is found that the electron ground-state
energy diminishes according to eq.~(2)
as $B$ increases within the Schwinger domain \hbox{$B \ll B_Q$,} 
but it reaches 
a {\it minimum value\/} of  $(1 - 4.6 \times 10^{-5}) m_e$  at $B = 0.25 \, B_Q$
and then rises \cite{cons72,gepr94}. 
At $B > 0.65 \, B_Q$ 
the electron grows heavier than $m_e$, but only slowly. The asymptotic
fractional enhancement, valid at very large $B$, is\cite{janc69}\hfil\break
\smallskip
\qquad \quad 
$(E_o - m_e)/m_e = (\alpha/ 4 \pi) \left( 
\left[\ln (2B/B_Q) - \xi - {3\over2} \right]^2 + \beta \right)$ 
\qquad $B\gg B_Q$ \hfil (3)\break 
to first order in $\alpha$, where $\xi = 0.577$ is Euler's constant, 
and $\beta\approx 3.9$ is a numerical constant (estimated here from the 
numerical integrations of ref.~\cite{cons72}).

  Thus, an electron's ground-state energy  is doubled, 
$E_o \sim 2 m_e$, only at $B \sim 10^{32}$ Gauss. \  (Higher-order
corrections might change this result somewhat.) Of course, the 
maximum fields 
attained in neutron stars fall far short of this. The dynamical
saturation field for convective motions in nascent neutron stars
is $\sim 10^{16}$ G;  
and $B\sim 3 \times 10^{17}$ G is possible if the free energy of 
differential rotation in a rapidly-rotating, newborn neutron star is 
efficiently converted by a post-collapse dynamo\cite{dt92,td93}. 
But if \hbox{$B> (8 \pi P Y_e)^{1/2} \sim 10^{17}$ G}, where $P$ is the 
pressure and $Y_e$ the electron fraction in the liquid interior of a 
neutron star, then buoyancy overcomes stable stratification 
and an inhomogeneous field is dynamically lost\cite{gold92,td93}.

For $B \sim 10^{17}$ G, eq.~(3) implies $E_o - m_e \approx 0.03 \,
m_e$.  Thus, magnetic self-energy corrections for electrons and
positrons are probably not important over the range of magnetic fields
and at the level of accuracy typically attained in neutron star
astrophysics.

\section*{Atoms and Molecules at \boldmath $B> B_Q$}

At sufficiently low temperatures, a magnetar's surface will be covered
with atoms and molecules.  This surface structure can have 
consequences for the star's quiescent X-ray emissions,
because it determines the work function for removing charged particles from 
the surface, as necessary for maintaining currents in the 
magnetosphere\cite{thom00}.
Such currents may result from magnetically-driven crustal deformations 
such as {\it twists\/} of circular patches of the crust.\footnote{Crustal 
twists, with spiral patterns of shear strain, may be a common type 
of magnetically-driven deformation. The
pressure in the crust is due mostly to degenerate particles (relativistic 
electrons, and free neutrons at densities above neutron drip), but 
the shear modulus is due only to relatively weak Coulomb forces of
the lattice. Hence 
the crust is relatively incompressible, and pure shear deformations 
allow the largest range of motion, with the greatest energy transfer 
between the crust and the magnetic field.}
If a bundle of field lines, describing an arch in the magnetosphere, has one 
footpoint twisted (with the motion driven from below by the evolving field), 
then a current must flow along the arch to maintain the twisted exterior
field, since $\oint {\bf B \cdot d\ell} = 4 \pi I/c$.  
Surface impacts of the flowing charges create hot spots at 
the arch's footpoints and ultimately
dissipate the exterior magnetic energy of the twist, with implications 
for SGR and AXP X-ray light curves and their time-variations\cite{thom00}.  
Here we focus on the atomic and molecular physics that comes into
play, following a paper by Ruderman \cite{rude74} and extending the 
arguments to $B>B_Q$.

  The Bohr radius of a hydrogen atom is  $r_o = \lambda_e/\alpha$. 
The quantum gyration radius, 
$r_{\rm gyr} = \lambda_e \, (B/B_Q)^{-1/2}$, 
is smaller than $r_o$ for $B> \alpha^2 \, B_Q = 2.4 \times 10^9$ G.  
This is the characteristic field strength at which 
magnetism radically alters the atomic structure of matter.\footnote{The largest 
field you are ever likely to encounter personally is $\sim 10^4$ G if you have 
an medical MRI scan. Fields $\gtrsim 10^9$ G would be instantly lethal.} 
At $B> \alpha^2 B_Q$, an atomic electron is constrained to gyrate along a 
cylinder which lies entirely within the spherical volume that 
the unmagnetized atom would occupy.  Electrostatic attraction binds the 
electron strongly to the central nucleus.
At $B\gg \alpha^2 \, B_Q$ the cylinder becomes very long and narrow, and 
atomic binding energies are adequately given by eigenvalues of the 
one-dimensional
Schr\"odinger equation.  A simple, intuitive estimate---which gives 
a good estimate of the ground state energy despite its lack of 
rigor---involves idealizing the atom as a line-charge of length $2 \ell$.
For linear charge density $e/2\ell$, the electrostatic energy is
$\varepsilon = - (e^2/ \ell) \ \ln [\ell/ \, r_{\rm gyr}]$.
A lower cutoff $r_{\rm gyr}$ is necessary because the charge distribution 
does not resemble a line when you get within $\sim r_{\rm gyr}$ of the nucleus. 
It is more like a sphere, contributing an energy $\sim - qe/r_{\rm gyr}$
where $q =  e r_{\rm gyr}/\ell$; but this contribution can be neglected in 
the limit 
$\ell \gg r_{\rm gyr}$ or $B \gg \alpha^2 B_Q$.  Thus, the ground state 
energy, including the energy of non-relativistic motion parallel to 
${\bf B}$, is
${\cal E}_o(\ell) = (\hbar^2 /2 m_e \ell^2) -  
  (e^2/ \ell) \, \ln [\ell/ \, r_{\rm gyr}]$.
Minimizing this according to $d {\cal E}_o/ d\ell = 0$, we find 
$\ell \simeq r_o \,[ \, \ln (r_o / r_{\rm gyr})\, ]^{-1}$. 
This shows that the {\it length\/} of the thin cylindrical atom is less than 
the Bohr diameter, but only by a modest, logarithmic factor. The ground state
\smallskip
hydrogen binding energy is then\hfil\break
\smallskip
\qquad\qquad\qquad $ {\cal E}_o \simeq - (\epsilon_o / 4) \ 
[ \, \ln (B/ \alpha^2 \, B_Q) \, ]^2 
\ \ \ \ \ \ \hbox{for} \ \ \ \ \ \ B\gg \alpha^2 B_Q,$ \hfil (4)\break
where $\epsilon_o = \alpha^2 m_e / 2 = 13.6$ eV is one Rydberg.  
Note that $E \propto [ \ln B ]^2$ energy scalings are ubiquitous in 
ultra-magnetized systems (cf.~eqs.~3,4,5). 

  As $B$ increases beyond $B\sim B_Q$, the radius of the atomic cylinder 
shrinks to less than the Compton wavelength 
but eq.~(4) remains a reasonably good approximation. 
This is because the electron's inertia for longitudinal motion 
($\parallel {\bf B}$) stays close to $m_e$ in the ground-state 
Landau level even at $B > B_Q$. \ 
\ Equation (4) would become invalid
if the longitudinal motion became relativistic.
But this would require  $\ell < \lambda_e$, which occurs only at
$B > \alpha^2 \, \hbox{exp}(2/\alpha) \, B_Q \approx 10^{115}$ G. \
Magnetic fields can never get this strong.  We will show that 
the vacuum breaks down at smaller $B$. 

Equation (4) implies that the binding energy of hydrogen 
near the surface of
a magnetar with $B \simeq 10 \, B_Q$, is ${\cal E}_o \simeq 0.5$ keV.  This is
comparable to the surface temperatures of some young magnetar 
candidates \cite{td96,heyl97}.  

There are two classes of {\it hydrogenic  excitations.}  
Longitudinal excited states are well-approximated as multi-nodal 
eigenfunctions of the 1-D Schr\"odinger equation; e.g., the first excited 
state has a node at the position of the 
nucleus. Transverse excited states involve transverse displacements 
of the center of electron gyration away from the nucleus. Semi-classically, 
the electron then experiences ${\bf E \times B}$ drift, and its center
of gyration moves in a circular orbit around the nucleus. (See 
ref.~\cite{rude74} for details.) Of course, Landau-level excitations 
are also possible, but the excitation energy is enormous for $B>B_Q$.
 Atoms generally become unbound when such free energies are present.

Longitudinal excitations tend to require more energy than transverse, 
so in ultra-magnetized {\it multi-electron atoms,} orbitals corresponding
to transverse hydrogenic states fill up before longitudinal.  
In fact, for $B >  Z^3 \alpha^2 B_Q \approx (Z/26)^3 \, B_Q$, 
where $Z$ is the electron number,
no orbitals with longitudinal nodes are occupied, and the atomic
structure is very simple \cite{rude74,lieb92}.  
Note that Fe$^{56}$, which is likely to be the
dominant nuclear species on a clean neutron star surface, has $Z = 26$.
Thus, this condition is satisfied on magnetars, but not on radio pulsars
with fields $\sim 10^{12}$ G. \  The atomic binding \smallskip
energy is then 
\hfil\break
\smallskip
\qquad\qquad\qquad\qquad ${\cal E}_o(Z) \simeq - (7/24) \ Z^3 \ \epsilon_o \
[ \, \ln (B/ Z^3 \, \alpha^2 \, B_Q) \, ]^2.$ \hfil (5)\break
\hbox{ \ \ \ When} $B \gg  Z^3 \alpha^2 B_Q \approx (Z/26)^3 \, B_Q$, 
atoms on a neutron star's surface form long polymer-like
molecular chains parallel to ${\bf B}$, bound by the electrostatic 
attraction of shared electrons.  The molecular binding energy per 
\smallskip
nucleus is \cite{rude74,neuh87,lai92} \hfil\break
\smallskip
\qquad\qquad\qquad\qquad\quad 
$\Delta {\cal E} \simeq - (3/2) \, Z^3 \, \epsilon_o  \
( B/ Z^3 \, \alpha^2 \, B_Q)^{0.37}.$\hfil (6)\break
Together, these results determine the approximate work function for ionic 
emission from a magnetar's surface\cite{thom00}.

\vfill
\eject

\section*{Vacuum Polarization and Radiative Processes}

Photon modes in the magnetized vacuum include the extraordinary
mode or \hbox{\it E-mode\/}, with 
oscillating electric vector  ${\bf E_E \perp B}$, and the ordinary
mode or \hbox{\it O-mode\/}, 
with \hbox{${\bf E_O \perp E_E}$}. \ Both electric vectors are also 
orthogonal to ${\bf k}$, the direction of propagation.\footnote{Photon 
eigenmodes are linearly polarized, as described here, except in narrow zones 
of ${\bf k}$-space where the angle between ${\bf k}$ and $\pm{\bf B}$ 
satisfies $\theta_{kB} \lesssim (\omega/m_e)^{1/2} \, (B/B_Q)^{-1/2}$
and $\hbar \omega$ is the photon energy.  \ 
For propagation along $\pm{\bf B}$ within these zones, the E and O modes are 
elliptically polarized; and circularly polarized for ${\bf k \parallel \pm B}$.}
Due to the process shown in Fig.~3, where the double-lines 
are propagators for a magnetized, virtual $e^+ \ e^-$ pair,
the indices of refraction of the two modes
are very different at $B>B_Q$:$\,$\footnote{The affect of a 
magnetized {\it plasma\/} on the eigenmodes and indices of refraction is
small in comparison to the magnetic vacuum polarization so long as
$\omega\gg\omega_{c2} \equiv (3\pi/\alpha)^{1/2} \ (B/B_Q)^{-1/2} \ \omega_p$
for $B\gg B_Q$, where $\omega_p = (4\pi N_e e^2/ m_e)^{1/2}$ is the plasma
frequency and $N_e$ is the electron density (see \cite{buli97} and 
references cited therein). This is satisfied in many or most applications to
observable phenomena in magnetar magnetospheres since
$\omega_{c2} = 0.13 \ (N_e/10^{23} \, \hbox{cm}^{-3})^{1/2} \
(B/10 \, B_Q)^{-1/2}$ keV.} 
\hfil\break  
\centerline{$n_{\scriptscriptstyle O} = 1 + (\alpha/6\pi) \ 
\sin^2 \theta_{kB} \ (B/B_Q)$}
\centerline{$n_{\scriptscriptstyle E} = 1 + (\alpha/6\pi) \ 
\sin^2 \theta_{kB}$. \ \ \ \ }

\begin{figure}[b!] 
\centerline{\epsfig{file=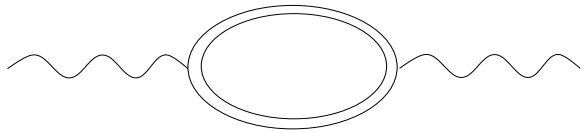,height=0.66in,width=3.02in}}
\vspace{10pt}
\caption{Vacuum polarization diagram.}
\label{fig3}
\end{figure}

If $n_{\scriptscriptstyle O} - n_{\scriptscriptstyle E} \gtrsim 
(k \ell_B)^{-1}$, where $k$ is the wavenumber and 
$\ell_B$ is the scale-length of variation of the magnetic field,
then the modes adiabatically track: E stays E and O stays O as photons 
move through the changing field geometry.
This condition is generally satisfied for X-rays in the 
magnetospheres of magnetars.
Shaviv, Heyl \& Lithwick \cite{shav99} used geometrical optics to
model {\it magnetic lensing\/} in the vicinity of a magnetar with a 
pure dipole field and a uniformly bright surface.  They found O-mode 
image distortion and amplification, varying with viewing
angle.  This remarkable effect may be hard to 
observe in practice because  fields
\hbox{$B\gtrsim (6 \pi/\alpha) B_Q\sim 10^{17}$ G} are required to produce 
strong lensing effects. Observations may also be complicated by
non-uniform surface brightness, gravitational lensing\cite{page95}, 
higher-order magnetic multipoles, photon splitting 
(see below) and X-ray emission from a magnetar's diffuse, Alfv\'en-heated halo.

When the excitation energy of the first Landau-level is much greater
than the photon energy,  
$\omega_B(1) \equiv m_e[(1 + 2B/B_Q)^{1/2} - 1] \gg \omega$,
then {\it photon scattering off electrons\/} is strongly suppressed in
the E-mode.
Semi-classically, this is easy to understand: the radiation electric field
(${\bf E_E \perp B}$) is unable to significantly drive electron
recoil.  Paczy\'nski first noted\cite{pacz92} that this 
greatly accelerates X-ray diffusion in the vicinity of magnetars, 
facilitating 
hyper-Eddington burst and flare emissions.  The E-mode scattering 
cross section, relative to Thomson, is 
$\sigma(E)/\sigma_T \sim (\omega/m_e)^2 \, (B/B_Q)^{-2}$ in the regime
of possible relevance for soft gamma repeater (SGR) bursts; 
see \S 3.1 of ref.~\cite{td95} for more details.

{\it Photon splitting and merging\/}, another important radiative effect, is 
depicted in Figure 4, with time advancing from left
to right for splitting, and right to left for merging.   These processes
are kinematically forbidden in free space, but they operate at $B > B_Q$ 
because the field acts as an efficient sink of momentum.
(Note the double-line, {\it magnetized\/} $e^-$ and $e^+$ propagators in Fig.~4.) 
The dominant splitting channel is $ E \rightarrow O \ \, O$. \ The rate 
\smallskip
for $B> B_Q$ and $\omega < m_e$ is\cite{adle71,td93b} \hfil\break
\smallskip
\qquad\qquad\qquad \qquad$\Gamma_{\rm sp} = (\alpha^3/2160 \pi^2) \ 
\sin^6 \theta_{kB} \ (\omega/m_e)^5 \ m_e$. \hfil (7)\break
(Splitting $E \rightarrow O \ E$ also occurs, but at a lower rate.)
Note that $\Gamma_{\rm sp}$ increases steeply with increasing photon energy; 
but it is independent of
$B$ for $B> B_Q$.  \ At $B < B_Q$ the process shuts down abruptly: 
$\Gamma_{\rm sp} \propto (B/B_Q)^6$.  \  

 How does this process affect SGR burst spectra? \
Simple splitting cascade models\cite{bari95} are illustrative but not 
realistic since O-mode photons do not split. Realistically, one must
consider the subtle interplay of splitting/merging 
and Compton scattering\cite{td95}.  In particular, E-mode splitting  
outside the E-mode scattering
photosphere produces O-mode photons which are isotropized by rapid Compton
scattering.  Subsequent mergers $O \, O \rightarrow E$ yield a quasi-isotropic
E-mode source function.  Only at $B < B_Q$ and outside the O-mode photosphere
do the modes truly decouple and all photons stream outward;
see \S 6 of ref.~\cite{td95} for many more details. 

\begin{figure}[b!] 
\centerline{\epsfig{file=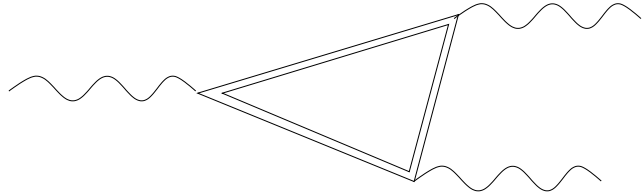,height=1.0in,width=3.33in}}
\vspace{10pt}
\caption{Photon splitting and merging in a strong magnetic field.}
\label{fig4}
\end{figure}

\section*{The Ultra-magnetized Pair Gas}

Ultra-strong magnetic fields also have profound {\it thermodynamic\/} effects.  
The magnetized photon-pair gas gives an example.  Such a gas may be 
created during an SGR burst or flare,  when a crust fracture or other 
magnetically-driven instability suddenly injects a large quantity of 
energy into the magnetosphere\cite{td95}.  The result is 
an optically-thick {\it trapped fireball\/}, 
confined by closed field lines, anchored to the star's surface. 
The gas inside this fireball has remarkable properties, 
as illustrated in Fig.~5. (This figure is included here courtesy of A. 
Kudari\cite{kuda96}.) The figure shows the ratio 
of pair energy density to the photon energy density, 
$\Lambda \equiv U_{e^+e^-}/ U_\gamma$, as a function of $T$ and $B$.
\ For $T \gg m_e$ and $T \gg \omega_B(1)$,
the magnetic field has little effect on the ultra-relativistic pairs:
\hbox{$U_{e^+e^-} = 2 \cdot (7/8) aT^4$},  so $\Lambda = (7/4)$. 
This should hold true across the whole right-hand side of Fig.~5, but
only 1000 Landau levels were used in making this graph, so the ratio
falls artificially below (7/4) at high $T$ and low $B$. 

\begin{figure}[b!] 
\centerline{\epsfig{file=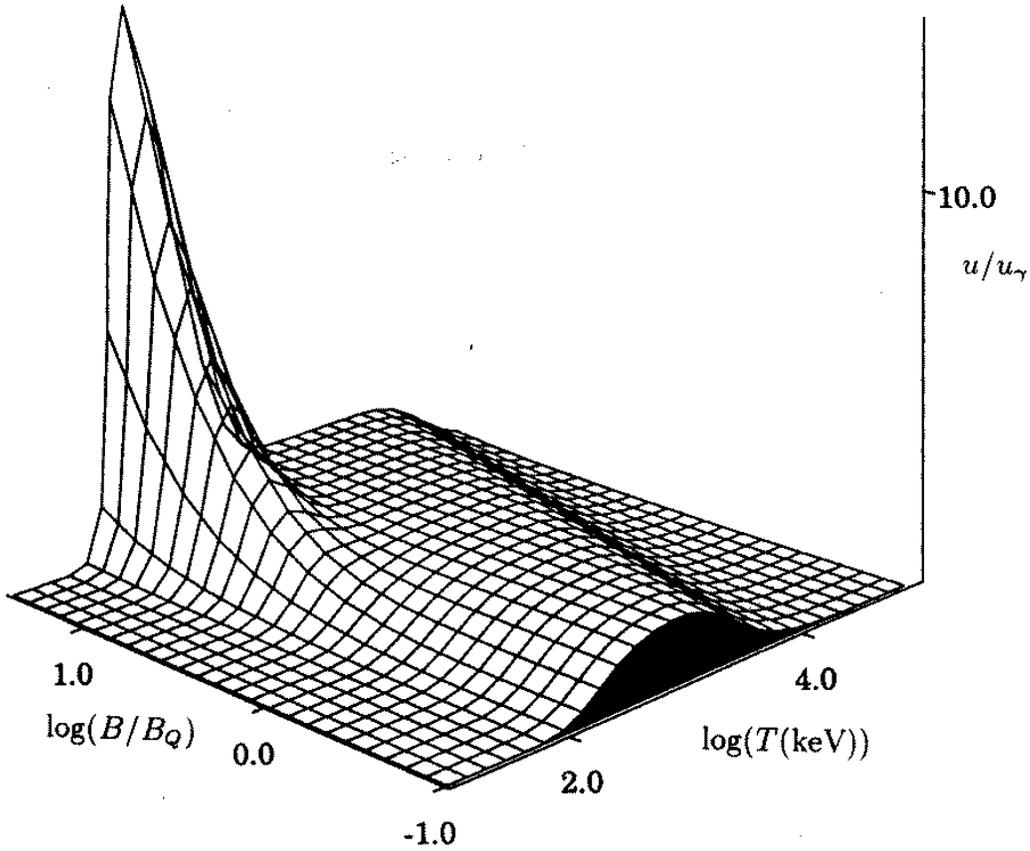,height=4.5in}}
\vspace{10pt}
\caption{The ratio of pair energy density to photon energy density in
a pair-photon gas, as a function of temperature and magnetic field 
strength.} 
\label{fig5}
\end{figure}

The striking peak in Fig.~5 is real, however.  It occurs for pairs with 
non-relativistic longitudinal motion, 
$T\ll m_e$, and $B>B_Q$. \  In this regime, only the first Landau level
is occupied: $T \ll \omega_B(1)$. \ The peak occurs because 
electrons and positrons are strongly localized in directions transverse 
to the field: $r_{\rm gyr} = \lambda_e \, (B/B_Q)^{-1/2}$. \ This allows
more of them to be packed into a given volume of ultra-magnetized gas.   
Formulae for thermodynamic parameters of a pair-photon gas in various limits 
are given in ref.~\cite{kuda96} and in \S 3.3 of ref.~\cite{td95}.  

Note that the trapped fireball of a common 
SGR burst, with energy $\Delta E \sim 10^{41}$ ergs confined within a
volume of order $(\Delta R)^3 \sim (10 \, \hbox{km})^3$ at $B\sim 10 \, B_Q$
has a temperature $T\sim 160$ keV \cite{td95,kuda96}.  This puts 
it right on the peak in Fig.~5$\,$! 
\vfill
\eject

\section*{Magnetic Vacuum Breakdown}

We argued above that a uniformly magnetized vacuum is stable 
against spontaneous electron-positron pair production.  Nevertheless, 
at sufficiently high $B$ the vacuum must break down.  Magnetic monopoles 
with mass $m_\eta$ and magnetic charge $\eta$ are spontaneously created when 
the energy they acquire in falling across a monopole 
Compton wavelength, $\varepsilon\sim \eta B \cdot (\hbar/m_\eta c)$, exceeds
their rest energy $m_\eta c^2$.  Dirac showed that a monopole charge
is an integral multiple of $\eta = (\hbar c/2e)$ from the condition that 
an electron wavefunction must be single-valued in the field of a 
monopole \cite{dira31}.
Thus, magnetic fields can never get stronger than \hfil\break
\smallskip
\qquad\qquad\qquad\qquad\qquad\qquad
$B_{\rm max} \sim \alpha \ (m_\eta/ m_e)^2 \ B_Q$ \ . \hfil (8)\break
A firm upper bound is $B_{max} \simeq 10^{55}$ G for Planck-mass monopoles, 
$m_\eta = 10^{19}$ GeV.  
GUT theories predict $m_\eta = 10^{16}$
GeV or $B_{\rm max} \simeq 10^{49}$ G. Superstring/M-theory predicts
intermediate values: 
$m_\eta = \alpha_s^{-1/2} = 10^{17}$--$10^{18}$ GeV, where 
$\alpha_s$ is the string tension, thus  \hfil\break
\smallskip
\qquad\qquad\qquad\qquad\qquad\qquad
$B_{\rm max} \simeq 10^{51} - 10^{53} \ \hbox{G}. $ \hfil (9)\break
\hbox{\ \ \ New} work shows that the energy scale for quantum gravity could 
be as low as 
\hbox{$M_o\sim 1$ TeV} if there exist ``large" extra dimensions to space
\cite{arka98}.  The extra dimensions are wrapped in 
closed geometries (e.g., circles) of size 
$L \sim  1 (M_o/1 \, \hbox{TeV})^{-2}$ millimeter for two extra dimensions,
or $L \sim \ell_p (M_o/M_p)^{-(n+2)/n} \, \ell_p$ for $n$
extra dimensions; where $M_p$ the Planck mass and $\ell_p$ is the 
Planck length. This would imply a small limiting field strength: 
$B_{\rm max} \simeq 10^{23} \, (m_\eta/1 \, \hbox{TeV})^2$ G. \  However,
there is no experimental evidence for large extra dimensions at the 
present time.  The most plausible upper limit is given by eq.~(9). 

Thus, {\it a vast range of tremendous field strengths are possible in 
Nature.} We don't yet know any objects that generate such fields,
but some possibilities have been suggested.
For example, superconducting cosmic strings---if they exist---could 
generate fields $\gtrsim 10^{30}$ G in their vicinities\cite{ostr87}. 
Perhaps future astrophysicists will regard neutron star magnetic
fields as mild$\,$!  

\section*{Magnetar Spindown}

  In this final section, I consider a topic of great current interest,
namely recent observations of soft gamma repeater spindown 
histories \cite{kouv98,hur99a,kouv99,mura99,mars99,wood99}, and their 
interpretation in the context of the magnetar model. 
At present, the most promising scenario involves
{\it episodic, wind-aided spindown\/} (\S 4 in ref.~\cite{thom00}).  
This is based upon several background developments.
In 1995 Thompson and I proposed that frequent, small-scale
fractures in the crust of a young magnetar produce quasi-steady seismic and 
magnetic vibrations, energizing the magnetosphere and driving a diffuse,
relativistic outflow of particles and Alfv\'en waves 
(\S 7.1.2 in ref.~\cite{td95}).  A year later we made a first estimate of this 
outflow's power \cite{td96}.  Thompson \& Blaes subsequently noted that 
a magnetar's rate of spindown is greatly 
accelerated by such a wind (\S VII B in ref.~\cite{tb98}).  All of 
this work pre-dated the discovery of X-ray pulsations from SGRs \cite{kouv98}. 

How strong is the wind?  \ The wind luminosity, $L_W$ scales roughly with
the magnetic energy density in the deep crust,\footnote{Most of 
the free energy in magnetars is stored in the {\it internal\/} magnetic field, 
probably in toroidal and high-multipole components, so $B_{\rm crust}$ 
is usually much greater than $B_{\rm dipole}$.} $\propto B_{\rm crust}^2$
\cite{td96}.  But if $B_{\rm crust} > (4\pi \mu)^{1/2}\sim 6\times 10^{15}$ 
G, where $\mu$ is the shear modulus in the deep crust, then
evolving magnetic stresses overwhelm lattice stresses 
and the crust deforms plastically instead of fracturing, choking 
off the Alfv\'en-powered wind.  This suggests an upper limit 
$L_W \lesssim 5 \times 10^{36}$ erg s$^{-1}$ for a $\sim 10^4$-year-old 
magnetar \cite{td96}.  In 1996, we proposed that a wind operating
near this upper limit could account for radio synchrotron nebula 
that seemed to surround SGR 1806$-$20 \cite{kulk94}. However,
we now know that the SGR is not coincident with this nebula \cite{hurl99}. 
There is no direct observational evidence for a quasi-steady
wind from any SGR. \ 
Magnetar winds must be mild enough to produce no detected radio 
emission, with $L_W$ probably much less than the theoretical upper 
limit of ref.~\cite{td96}, because this limiting value assumed 
optimal conditions, including the dubious application of a formula at the edge 
of the regime where it breaks down \hbox{(i.e., $B_{\rm crust} \sim B_\mu$).}  
It is likely that $L_W$ is 
comparable to the steady X-ray luminosity emitted by the hot stellar
surface and Alfv\'en-energized halo: $L_{W} \sim 10^{35}$--$10^{36}$ 
erg s$^{-1}$.  

Rothschild, Marsden and Lingenfelter have plotted 
two graphs, included in this volume, which nicely elucidate 
constraints on SGR spindown
for constant $L_W$ and $B_{dipole}$, based upon formulae derived independently 
in refs.~\cite{hard99,thom00}.\footnote{These references correct an 
inaccuracy in the original wind-aided spindown rate given by ref.~\cite{tb98}.} 
These plots show that wind
luminosities $L_W\lesssim 10^{36}$ erg s$^{-1}$ imply 
$B_{\rm dipole} \gtrsim 10^{14}$ G in order to match the observed 
values of $P$ and $\dot{P}$; but the implied stellar ages are then 
moderately shorter than the estimated ages of the putative associated supernova 
remnants (SNRs).  Of course, the SNR associations or ages may be 
unreliable, since the SGRs lie far from the SNR centers, and the ages
are only rough order-of-magnitude estimates.  However, we favor a different 
interpretation:  the wind 
is probably {\it episodic\/}, so the effective spindown age of the star is 
less than the SNR age.   In particular, we proposed (in \S 4.1--4.2 of
ref.~\cite{thom00}) that strong winds and rapid spindown prevail 
only during limited episodes of a young magnetar's life, when it 
is magnetically active and observable as an SGR.  This fits in nicely 
with observations of anomalous X-ray pulsars (AXPs).
These objects have spindown ages $P/2\dot{P}$ that are comparable
or {\it longer\/} than the ages of their associated SNRs\cite{gott99,kasp99}, 
suggestive of 
young magnetars observed during their non-windy, inactive episodes.  
A fully consistent scenario is possible \cite{thom00}. 

Note, incidentally, that $B_{\rm dipole} \lesssim B_Q$ is possible 
in a magnetar if the lowest-order magnetic moment decays quickly, 
e.g., via the Flowers-Ruderman instability\cite{fr77} (see \S 14.2 and 15.2 in
ref.~\cite{td93}; \S 7.1.2 in ref.~\cite{td95}).  This is because 
a magnetar is a {\it magnetically-powered\/} neutron star:  its emissions 
depend upon the {\it total free energy and configuration of its magnetic field, 
not simply upon its exterior dipole moment.}  The light curve 
of the 1998 August 27 giant flare gives evidence for strong higher-order 
multipole moments in SGR 1900+14\cite{fero00}. SGR bursts give evidence for 
magnetically-powered activity (e.g., 
refs.~\cite{td95,fero00,palm99,gogu99}). 

Marsden et al.~\cite{mars99} also suggested that the spindown rate of 
SGR1900+14 was enhanced 
by a factor $\sim 2$ during the summer of 1998. In the context 
of the magnetar model, this could mean that $L_W$ increased by a factor
$\sim 4$. \  Possible evidence for this comes from $\dot{P}$ measurements 
during RXTE runs immediately preceding and following the interval in
question\cite{kouv99};  but it should be noted 
that RXTE was observing the SGR at those times as 
a ``target of opportunity" because the star was emitting hundreds of 
bursts \cite{kouv99,wood99}.  
Transient accelerated spindown during episodes of vigorous bursting can 
occur in the magnetar model, because the relativistic outflow may be enhanced.  
But only a handful of bursts were detected by BATSE during mid-summer 
of 1998 \cite{wood99}, and the RXTE/ASCA determination of $\dot{P}$ 
between 1998 Aug.~28 and Sept.~17 was $6.2\times10^{-11}$ s/s\cite{wood99} 
(a number that was rounded up to $1.\times10^{-10}$ in ref.~\cite{mura99}).
Furthermore, spindown rates measured over short time intervals, such as during 
single RXTE runs, can be affected by other transient or periodic 
effects (e.g., free precession: see \S 4.3 in ref.~\cite{thom00};
also ref.~\cite{mela99}). 

Although an increase in $\dot{P}$ by $\sim 2$ during the summer of
1998 cannot be ruled out, we suspect that the average spindown rate 
was similar to that which prevailed at other times during the past few years, 
and the observed shift in the spindown history was due to an abrupt 
{\it spindown glitch\/} during the extraordinary giant flare of 1998 
August 27th.    Such a glitch could be caused by the unpinning of crustal 
superfluid vortices in a magnetar with a crust that has been deformed by 
evolving magnetic stresses\cite{thom00}.  

In conclusion, we have come a long way from the days of ref.~\cite{dt92}
when simple magnetic dipole radiation seemed to be an adequate idealization 
for SGR spindown! \ More observations are needed to determine 
whether glitches really occur in SGRs and AXPs \cite{td96,heyl99,thom00}
and what sign they may have; whether these stars
exhibit free precession (which could give us the first direct measure of a 
magnetar's {\it internal\/} field\cite{mela99,thom00}); and to further
test and constrain models of these fascinating stars.  

\bigskip

\noindent {\it Acknowledgments:\,} This work was supported by NASA grant
NAG5-8381 and Texas Advanced Research Project grant ARP-028.

\end{document}